\documentclass[showpacs,preprintnumbers,amsmath,amssymb,12pt,floatfix,epsfig]{revtex4-1}
\usepackage{graphicx}

\begin{document}
\title{Kinematics of a cascade decay
 for the precision measurement of $^{29}{\rm Si}$ binding energy}
\author{Yongkyu Ko\footnote{yongkyu.ko@gmail.com}, 
and K. S. Kim\footnote{Corresponding author: kyungsik@kau.ac.kr}}
\affiliation{Research Institute for Basic Sciences, Korea Aerospace
University, Koyang 412-791, Korea }

\date{\today}

\begin{abstract}
Comparison of a Penning trap and a flat-crystal spectrometer experiments
gives a direct test of $E=mc^2$. The result is $1-\Delta mc^2/E
=(-1.4 \pm 4.4) \times 10^{-7}$ for $^{29}{\rm Si}$ and $^{33}{\rm S}$. 
The dominant uncertainty is on the $\gamma$-ray measurement 
in neutron capture reactions, and the secondary $\gamma$-ray has 
the uncertainty  4.0 eV for $^{29}{\rm Si}$. We calculated the
Doppler effect of the secondary $\gamma$-ray as $-646.9 \cos \theta$
eV from the relativistic energy momentum relation of 
the $^{28}{\rm Si}(n,\gamma)^{29}{\rm Si}$ reaction.
This corresponds to the full wave of half maximum of 431.3 eV.
The error 4.0 eV comes mainly from
the Bragg angle measurement between the centroids of the linewidths
which means that only the most probable part 
of the whole data has been considered.
It is necessary to confirm the assumption of the isotropy for the object 
to measure. We discussed a coincidence measurement as 
one of the methods to overcome the assumption.

\end{abstract}


\maketitle


Nuclear binding energy is the mass difference between
initial and final states. The Penning trap in the recent report \cite{science,shi}
measures the masses with fractional uncertainty below $10^{-11}$.
Another method is to measure the $\gamma$-ray energy emitted
from the transition between initial and final states. The flat-crystal
spectrometer in the recent report \cite{dewey} measures 
the $\gamma$-ray energy
with relative uncertainty below $10^{-6}$. The comparison of the two 
methods is a direct test for Einstein's famous formula $ E=mc^2$, 
and the result is $1-\Delta mc^2/E=(-1.4 \pm4.4)\times 10^{-7}$ \cite{nature}.
In the report the systematic error on the
comparison is currently dominated by the uncertainty on the
$\gamma$-ray measurements.  

The binding energy is the sum of the $\gamma$-ray energy and 
the recoil energy in traditional $\gamma$-ray measurements and 
the flat-crystal measurement \cite{dewey}. 
This is the result of non-relativistic
energy momentum relation. From relativistic energy momentum
relation we calculated the Doppler effect of the secondary 
$\gamma$-ray due to the recoil ($39 km/s)$ of the intermediate
nucleus as $-646.9 \cos
\theta$ eV. This uncertainty is far greater than the error $4.0$ eV
of the secondary $\gamma$-ray energy of  $^{29}{\rm Si}$
reported in Ref. \cite{dewey}.
Increasing the number of measurements can reduce the
statistical error to  $431.3 \sqrt{2}/\sqrt{147}=50.3$ eV, 
but can not reduce to the reported value 4.0 eV 
which comes chiefly from the measurements of 
the Bragg angle between the centroids of 
the spectrum and the lattice spacing of the silicon crystal.
Thus, the error bar is shown to be quite out of the diagonal line
at the level of 1.2 $\sigma$ in Fig. 5 of Ref. \cite{dewey}.

A secondary $\gamma$-ray has angular correlation 
with respect to the primary $\gamma$-ray in
a cascade decay \cite{biedenharn,roy}.
We calculated the angular correlation function 
for the decay of  $^{29}{\rm Si}$.
The most probable directions of the secondary
$\gamma$-ray are parallel and anti-parallel with the direction of
the primary $\gamma$-ray. 
The Doppler shift of the secondary
$\gamma$-ray is maximum at these angles.  Therefore, the average
value of the two energies of the Doppler shifted $\gamma$-rays at
angles $0^0$ and $180^0$ is free from Doppler broadening caused by the
recoil of the nucleus. Thus, the profile of the shifted
$\gamma$-ray might be even closer to a natural linewidth which
can also give the lifetime of the intermediate state of the
nucleus. The Institut Laue-Lagnevin (ILL) high-flux reactor 
provides one of the most intense neutron sources in the world 
and has a through beam tube \cite{kessler}. This leads us
to calculate the kinematics of the cascade 
decay for $^{29}{\rm Si}$ nucleus.

\begin{figure}
\includegraphics[width=1.0\linewidth]{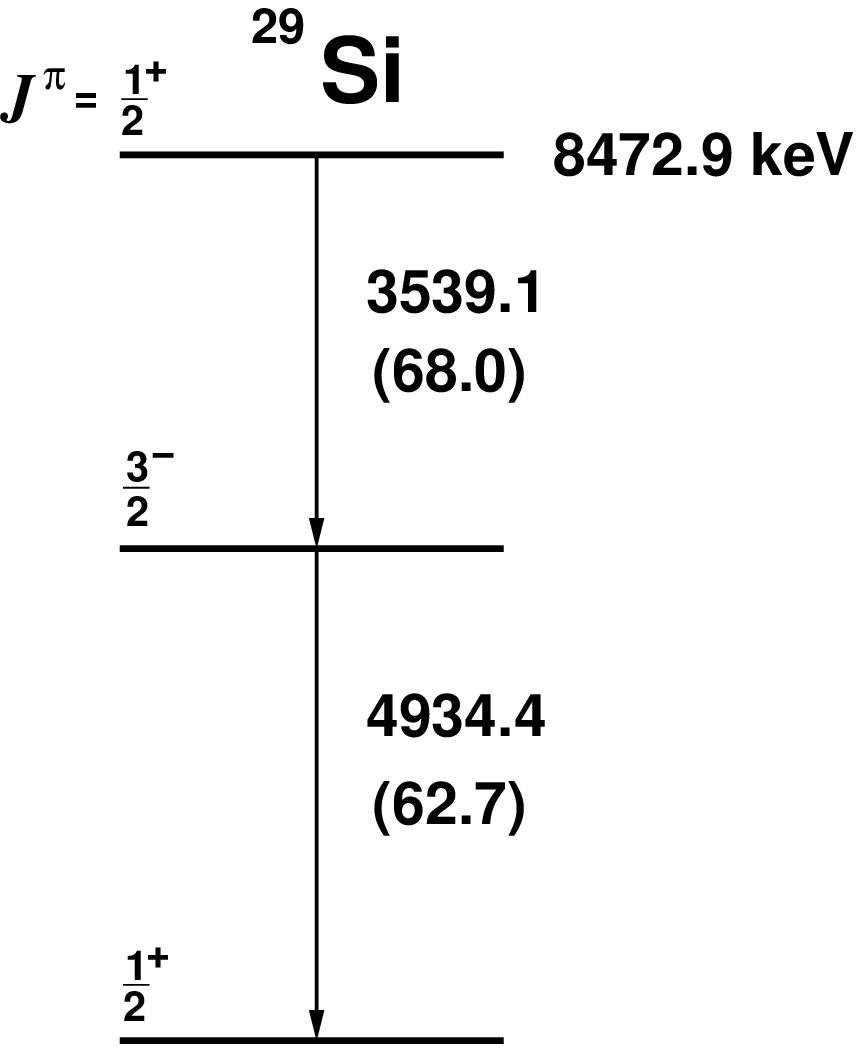}
\caption{Partial decay scheme for $^{29}{\rm Si}$.  The numbers in
parentheses are the number of $\gamma$-rays per 100 neutron
captures \cite{lone}. The spin and parity is referred to Ref.
\cite{raman}} \label{decayscheme}
\end{figure}

\begin{figure}
\includegraphics[width=1.0\linewidth]{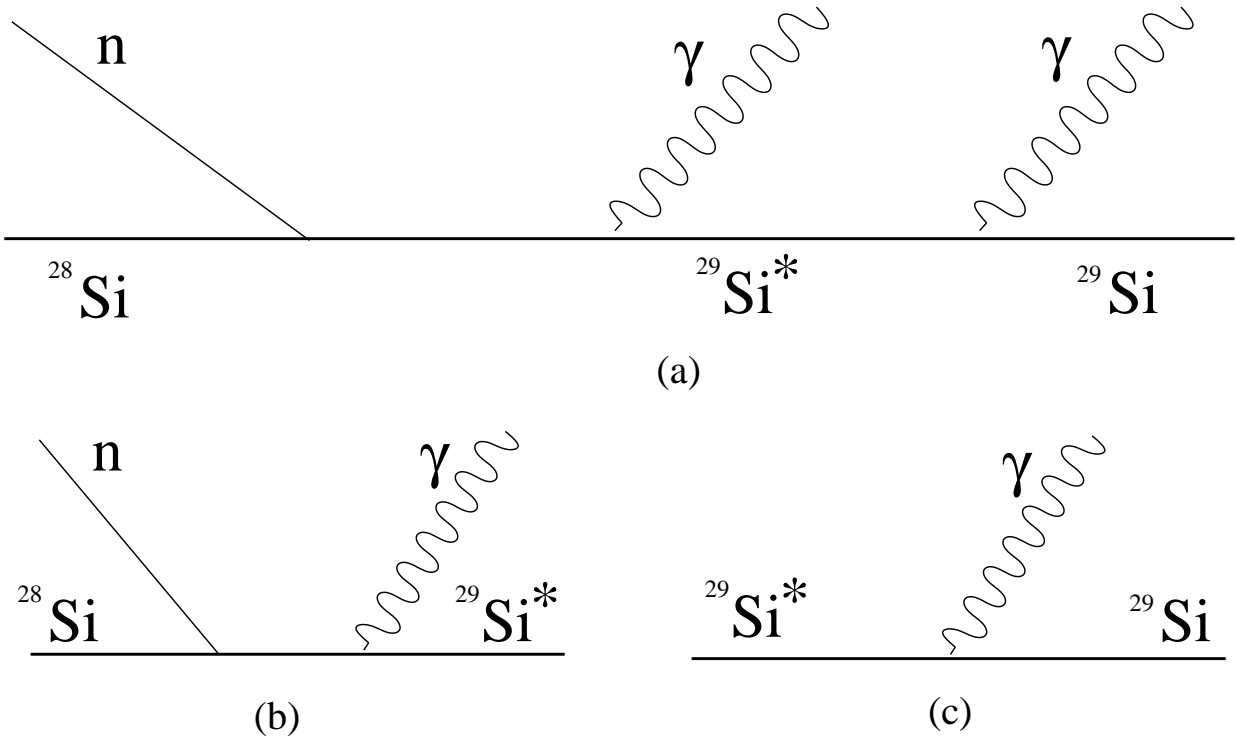}
\caption{(a) Feynman diagram of a neutron capture reaction
$^{28}{\rm Si}(n,\gamma, \gamma)^{29}{\rm Si}$.
(b) The first part of the neutron capture reaction
 can be regarded as a neutron
capture reaction $^{28}{\rm Si}(n,\gamma)^{29}{\rm Si}^*$ with
emitting a single $\gamma$-ray.
(c)  The second part of the neutron capture reaction
can be regarded as
a transition of an excited nucleus to the ground state with
radiating a $\gamma$-ray. } \label{decaysi29}
\end{figure}

The binding energy for a neutron capture reaction is the mass
difference between initial and final states as follows:
\begin{equation}
m(n)+m(^AX)=m(^{A+1}X)+S_n, \label{bind}
\end{equation}
where $m$ means the mass of the particle in parentheses, $A$ is an
atomic number, and $S_n$ is the separation energy of a neutron or
the binding energy of a neutron which is carried by the $\gamma$-rays
and the recoiled nuclei. If a single $\gamma$-ray is emitted dominantly 
in a neutron capture reaction, the problem has been considered 
in determining the deuteron binding energy \cite{ko}. In
this letter we are interested in a cascade decay emitting two
$\gamma$-rays successively which is the most probable channel. The
excited state of  $^{29}{\rm Si}$ is the case as shown in Fig.
\ref{decayscheme}, and the values in parentheses are the number of
$\gamma$-rays emitted per 100 neutron captures \cite{lone}.  This
decay channel can be represented as a Feynman diagram shown in
Fig. \ref{decaysi29} (a) and can be separated into two parts as Figs.
\ref{decaysi29} (b), and (c). The equation of the
binding energy Eq. (\ref{bind}) is also separated into two parts
corresponding to the separated Feynman diagrams, respectively, as
follows:
\begin{eqnarray}
m(n)+m(^AX) &=& m(^{A+1}X^*)+E_{b1}, \\
m(^{A+1}X^*) &=& m(^{A+1}X)+E_{b2}, \\
S_n &=& E_{b1}+E_{b2},
\end{eqnarray}
where asterisk means an excited state of the final nucleus.

Fig. \ref{decaysi29} (b) represents the neutron capture reaction
$^{28}{\rm Si}(n,\gamma)^{29}{\rm Si}^*$, and the energy-momentum
conservation law is given by
\begin{eqnarray}
m+M &=& E_1+\omega_1, \\
0 &=& \mbox{\boldmath{$p$}}_1+\mbox{\boldmath{$k$}}_1,
\end{eqnarray}
where $m=m(n)$ and $M=m(^{28}{\rm Si})$. The energy-momentum
relations of the final state are
$E_1^2=\mbox{\boldmath{$p$}}_1^2+M_1^2$ for the intermediate state
of the silicon nucleus and $\omega_1^2=\mbox{\boldmath{$k$}}_1^2$
for the primary photon. It should be noted that the velocity of
the center of mass for the neutron and the silicon system is only
114 $m/s$ for the incident neutron flux energy 0.056 eV, while
that for the neutron and the proton system is 1.6 $km/s$ as
calculated in Ref. \cite{ko}.  Hence, the Doppler effect of the
silicon nucleus due to the incident neutron is negligible ($-1.3
\cos \theta$ eV) so that it is not necessary to consider the
kinetic energy of the incident neutron \cite{dewey}.

Solving the equations of the energy-momentum conservation laws
with respect to the primary photon energy and replacing the
initial masses by the binding energy, one can obtain the binding
energy of the intermediate nucleus as follows:
\begin{equation}
E_{b1}=\omega_1-M_1+\sqrt{\omega_1^2+M_1^2}
\cong \omega_1+ {\omega_1^2 \over{2 M_1}}, \label{bind1}
\end{equation}
which is reduced to the non-relativistic result that is usually
used in the literature.  Since the incident kinetic energy of the
neutron is ignored, there is no Doppler effect in this equation.
The last term in the above equation is the kinetic energy of the
recoiling intermediate nucleus, and its velocity can be
checked as follows:
\begin{equation}
KE= {1\over2}M_1v_{int} ^2=
{\omega_1^2 \over{2 M_1}}, \mbox{    }  v_{int} \cong 39 km/s,
\end{equation}
where the natural unit($\hbar=c=1$) is used in every equation
throughout this letter, and thus, the velocity in SI unit should be
multiplied by the light velocity.

Since the velocity causes the secondary $\gamma$-ray to be
Doppler shifted considerably, it is important to take into account the
kinetic energy and momentum of the intermediate nucleus in
calculation. Hence, the law of the energy-momentum conservation for
the process shown in Fig. \ref{decaysi29} (c) should be given by
\begin{eqnarray}
E_1 &=& E_2+\omega_2, \\
\mbox{\boldmath{$p$}}_1 &=& \mbox{\boldmath{$p$}}_2
+\mbox{\boldmath{$k$}}_2,
\end{eqnarray}
where the energy-momentum relation of the ground state silicon
nucleus is $E_2^2=\mbox{\boldmath{$p$}}_2^2+M_2^2$, and that of
the secondary photon is $\omega_2^2=\mbox{\boldmath{$k$}}_2^2$. To
specify the angle between two photons that is of importance in the
Doppler effect and angular correlation, it is useful to show the
following intermediate step:
\begin{equation}
(E_1-\omega_2)^2=E_2^2=M_2^2+(\mbox{\boldmath{$p$}}_1^2+
\mbox{\boldmath{$k$}}_2^2-2p_1k_2 \cos \theta),
\end{equation}
where the direction of the primary photon is $\theta =\pi$, and the
direction of the recoiling nucleus is $\theta =0$. Solving the
equation as the same way of the previous energy-momentum law, one
can obtain the binding energy of the ground state of the silicon
as follows:
\begin{eqnarray}
E_{b2} &=& \omega_2-M_2+\sqrt{M_2^2+\omega_2^2
+2\omega_2 \{E_{b1}-\omega_1(1+\cos \theta) \} } \nonumber\\
&\cong& \omega_2+{\omega_2^2 \over{2 M_2}}
+{\omega_1^2 \omega_2 \over{2 M_1 M_2}}
-{\omega_1 \omega_2 \over{M_2}} \cos \theta, \label{bind2}
\end{eqnarray}
where the second term is the kinetic energy due to recoil, the
third term is negligible kinetic term, and the last term
represents the Doppler effect as shown in Ref. \cite{ko}. 
Higher order terms in expansion include the usual relativistic 
correction term which depends on the square of velocity.
It is negligible and is omitted \cite{moreh}. 
This result is calculated in laboratory frame and is consistent with
the result that is Lorentz transformed from the calculations in
rest frame of $^{29}{\rm Si}^*$.

The total binding energy of the silicon is the sum of Eqs.
(\ref{bind1}) and (\ref{bind2}) as follows:
\begin{eqnarray}
S_n &=& E_{b1}+E_{b2}\cong \omega_1+\omega_2
+{\omega_1^2 \over{2 M_1}}+{\omega_2^2 \over{2 M_2}}
+{\omega_1^2 \omega_2 \over{2 M_1 M_2}}
-{\omega_1 \omega_2 \over{M_2}} \cos \theta \nonumber\\
&=& 3538966.3(1.6)+4933946.3(4.0)+232.9
+450.9+0.0-646.9 \cos \theta \mbox{ eV}. \label{dop}
\end{eqnarray}
If one is able to handle the angle in the secondary $\gamma$-ray
measurement, the Doppler shifted $\gamma$-ray is detected.
Otherwise, all the shifted $\gamma$-rays reach the detector and
appear in the spectrum as a broadened lineshape. 
Thus, the 4934-keV profile is significantly Doppler broadened,
which decreases the accuracy with which the Bragg angle
can be determined \cite{dewey}.

\begin{figure}
\includegraphics[width=1.0\linewidth]{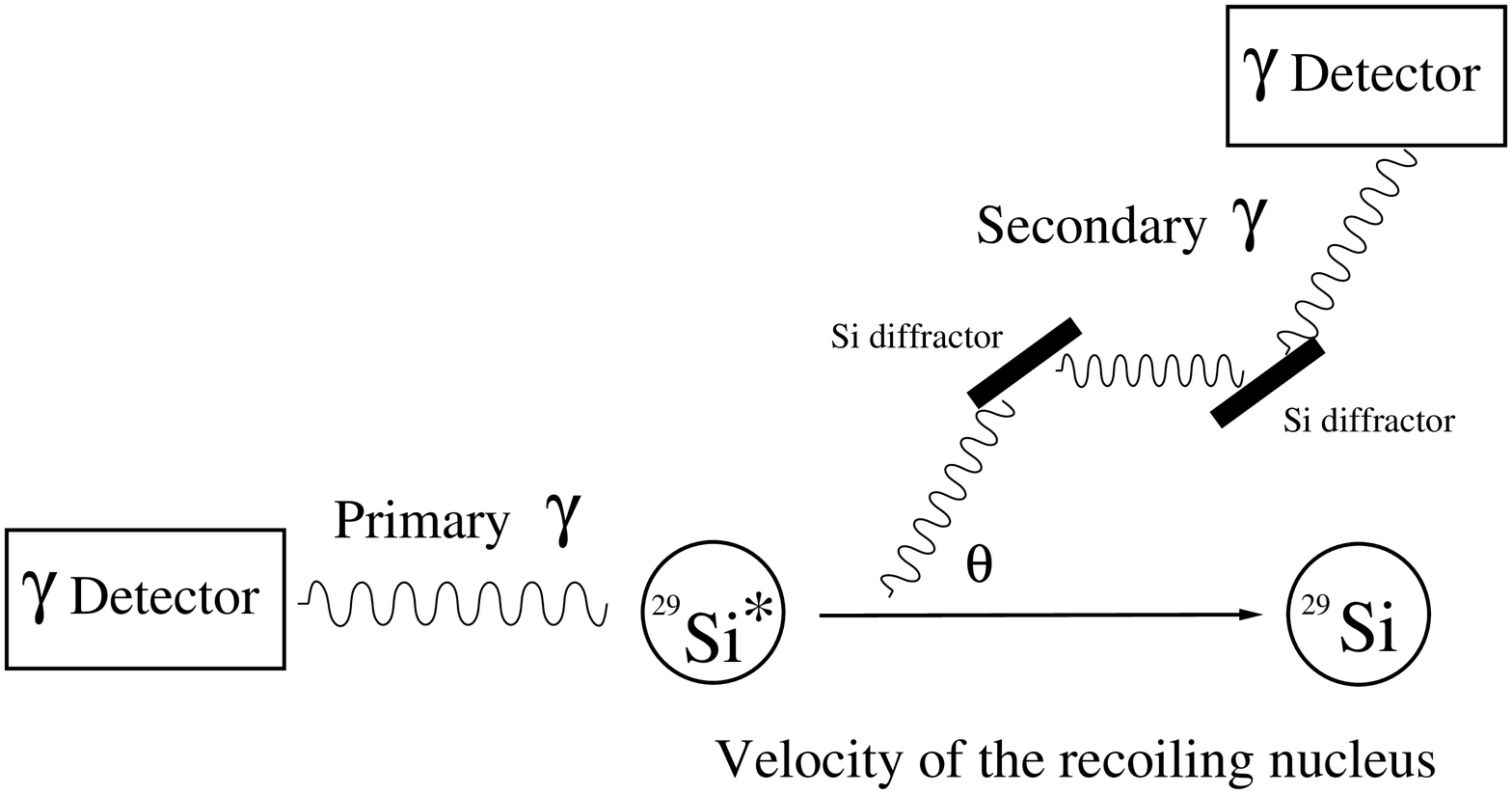}
\caption{Schematic diagram of a coincidence measurement with
considering the angular correlation of the two $\gamma$-rays.
The two $\gamma$ detectors check the coincidence of the $\gamma$-rays.
 } \label{angularcorrel}
\end{figure}

Assuming that the motion of the intermediate nucleus 
is completely isotropic and the Doppler effect causes nothing
in the central value, 
the authors of Ref. \cite{dewey} obtained the uncertainty 
of $4.0$ eV mainly from the measurements of the Bragg angle 
between the centroids of the diffracted $\gamma$-rays
by increasing the number of measurements up to 147 compared
to 42 for the primary $\gamma$-ray measurement. 
If one want to obtain the uncertainty from the linewidths
at the level of an uncertainty 4.0 eV,
the number of measurements should be more 
than $2(431.3/4.0)^2\approx 23252$ according to
the basic rule of the error propagation:
$\sigma_z=\sqrt{\sigma^2_x+\sigma^2_y}$ for $z=x-y$. 
The linewidth is too 
broad compared to the resolution of the flat-crystal 
spectrometer, but the measurement of Bragg angle between 
the centroids of the recorded profiles could reduce
the uncertainty to 4.0 eV by the assumption. 
This kind of an assumption about the
centroid measurement has a possibility to reduce 
the uncertainty to less than a natural linewidth,
and change the lifetime of the state from 
which  a $\gamma$-ray with a natural linewidth
is emitted. The assumption should be confirmed
by a proper method.
In a coincidence measurement of two successive emitting 
$\gamma$-rays, the secondary $\gamma$-ray is not broadened
but shifted. The magnitude of shifted energy depends on the angle 
between the primary and secondary $\gamma$-rays.
The probability distribution of the secondary $\gamma$-ray
with the angle is the angular correlation \cite{biedenharn}.
Fig. \ref{angularcorrel} shows a schematic diagram of a 
coincidence measurement. This is also an diffraction experiment
of the secondary $\gamma$-ray which is correlated with
the primary $\gamma$-ray and may help to understand
quantum physics: one photon is diffracted.

Since the polarizations of the $\gamma$-rays are not observed
and only directional correlation is observed,
the angular correlation function \cite{biedenharn,roy} is simply given by
\begin{equation}
W(\vartheta)= \sum_{l=even}^{l_{max}} A_{l}P_l(\cos \vartheta)\label{w},
\end{equation}
where $P_l$ is the Legendre polynomial of order $l$.  The
coefficient for the transition of angular momentum states $J_A
\rightarrow J_B \rightarrow J_C$ with emitting photons of the
angular momenta $L_1$ and $L_2$ in order
 is given by
\begin{eqnarray}
A_l &=& F_l(L_1,J_A,J_B)F_l(L_2,J_C,J_B), \\
F_l &=& W(J_BJ_AlL;LJ_B)C_{l0}(LL),
\end{eqnarray}
where $W(J_BJ_AlL;LJ_B)$ is a Racah coefficient, and $C_{l0}(LL)
\propto(-1)^{L-1}(2L+1)<L,L,1,-1|l0>$ is known as the radiation
parameter.   The angle of the binding energy in Eq. (\ref{dop}) is
related with the angle in the correlation function as $\theta =
\pi - \vartheta$, but it is not necessary to distinguish between
the angles, because only even power of the Legendre polynomial
contributes to the angular correlation function. For a simple
example, if a dipole-dipole transition occurs through the nuclear
states of $J_A=J_C=0$, and $J_B=1$, the angular correlation
function is $W(\vartheta)={3 \over{16 \pi}}(1+\cos^2 \vartheta)$.
If a dipole-dipole transition takes place through the nuclear
states of $J_A=J_C=1/2$ , and $J_B=3/2$ as assigned from the
$\gamma$-rays measurement in Ref. \cite{raman} and shown in Fig.
\ref{decayscheme},  the angular correlation function is
\begin{equation}
W(\vartheta)={3 \over{32 \pi}}(1+{3 \over7} \cos^2 \vartheta).
\end{equation}
One can see here that the secondary $\gamma$-ray has the most
probable distribution at the angles $\theta = 0$, or $\pi$.
The Doppler shift is maximum at these angles in Eq. (\ref{dop}). 

However, the surrounding environment in
which the recoiling nucleus is located is not a free space.
It is not sure that the shifted $\gamma$-ray at the
angle $\theta=0$ gives exact information on the binding energy.
Moreover, the surrounding environment varies from nucleus to
nucleus.  For example, the velocities of the intermediate state
of other nuclei $^{33}{\rm S}$ and $^{36}{\rm Cl}$
are greater than that of $^{29}{\rm Si}$, 
but they suffer from the Doppler broadening 
less than the case of $^{29}{\rm Si}$ \cite{dewey}. 
The reason can be that the
lifetime of the intermediate state of $^{33}{\rm S}$ and
$^{36}{\rm Cl}$ is much longer than that of $^{29}{\rm Si}$ as
shown from experimental data \cite{endt,skorka}. 
The velocity of these
intermediate nuclei could be attenuated during their long
lifetimes, and thus, the Doppler broadening would be considerably
reduced.  This aspect also reflects in the branching ratio of the
decay, that is, the number of $\gamma$-ray per 100 neutron capture
in Fig. 1 of Ref. \cite{dewey}. Since the lifetime of the
$^{29}{\rm Si}$ intermediate nucleus is quite short, it is
likely that the velocity of the intermediate nucleus is a little
slowed down, and the angular correlation between the two
successive $\gamma$-rays is clear. 

Let us think twice coincidence measurements 
at angle $\theta=0$ and angle $\theta=\pi$. 
Using Eq. (\ref{dop}), one can obtain the following
quantities:
\begin{eqnarray}
\bar{S}_n&=&{1 \over2} \{ S_n(\theta=0)+ S_n(\theta=\pi) \}, \label{avgbe}\\
\Delta E&=& {1 \over2} \{S_n(\theta=0)- S_n(\theta=\pi)  \} 
\cong \omega_2 \beta F \label{atten},
\end{eqnarray}
where the first equation is the desired binding energy, and the
second equation contains the magnitude of the Doppler shift attenuated
by the surrounding environment, namely, attenuation factor $F$ \cite{lieb,borner}. 
$\beta$ is the velocity of the intermediate nucleus.
The average binding energy $\bar{S}_n$ is free from
the major uncertainty due to the Doppler effect of the
secondary $\gamma$-ray.
Since the primary and secondary $\gamma$-rays
come from the electric dipole transitions $E_1$ in view of
the spin-parity of the states shown in Fig. \ref{decayscheme},
a coincidence measurement may rule out uncertainties
due to other multipole transitions.

The lifetime of the 4943 keV in old data \cite{endt,skorka} 
is 1.16 fs which corresponds to the linewidth of  0.57 eV. 
The resolution of the flat-crystal spectrometer is this order. 
Hence, it may be possible to extract the lifetime 
from the thermal broadened linewidth which can
be obtained from the classical Maxwell-Boltzmann distribution by
virtue of the modified Lorentzian shape used in Ref. \cite{borner}.
This can be compared with the Doppler shift attenuation
method.
If the energy difference Eq. (\ref{atten}) is compared to
the calculated value in Eq. (\ref{dop}), one can extract the
attenuated velocity of the recoiling nucleus which provides a
lifetime in the Doppler shift attenuation method.  

In conclusion, the equipment of a flat-crystal spectrometer is
an extremely accurate tool for a $\gamma$-ray measurement.
The recorded profiles along with interferometer angles 
have valuable physical facts such as the influence of
Si diffractors,
a Doppler broadened linewidth, an attenuated linewidth, 
or a thermal broadened linewidth. 
For example, the 6111 keV 
profile of $^{36}{\rm Cl}$
in Fig. 3 of Ref. \cite{dewey} shows the linewidth of
93.5 eV for $\Delta \theta=0.01$ arcsec where a
scale is 0.05 arcsec and its uncertainty 4.0 eV 
corresponds to 0.00043 arcsec. 
Likewise, the average linewidth of the recorded profiles,
especially $^{29}{\rm Si}$, should have been 
shown for the justification of the accuracy 4.0 eV 
in Ref. \cite{dewey}. It is insufficient to understand the
accuracy 4.0 eV from the broadened width 431.3 eV
without further information on the first observed data.
Usually, centroid measurements have been made in the past when
the resolution of a detector is greater than the linewidth,
and the assumption for symmetry have been tried
to prove \cite{green,wood}. Moreover, the assumption is not for
the object to measure but for the detector.
An assumption for the object to measure is
unusual in any experiment. As a result,
the 1.2 $\sigma$ discrepancy is an evidence for 
observing only part of the whole data.

A flat-crystal spectrometer is more suitable to measure 
sharp $\gamma$-rays or natural linewidths without
unnecessary assumptions rather than 
to measure broadened $\gamma$-rays
due to the Doppler effect. 
A coincidence measurement of two
$\gamma$-rays with considering their angular correlation can
reduce the uncertainty caused by the Doppler effect 
in the determination of binding energy.  This
improved binding energy measurement would lead 
to one step upgrade, at least 1.2$\sigma$ discrepancy,
for the direct test of $E=mc^2$ and relativistically consistent
comparison between energy and mass \cite{nature}. 
If the measurement
technique for cascade decay is developed, one can confirm
one of the postulates of quantum physics and learn the more
accurate information about the nuclear structure, for examples,
accurate energy level, the lifetime of the nuclear intermediate
state, and so on.

\section*{Acknowledgements}
This work was supported by the National Research Foundation 
of Korea (Grant No. 2009-0074238).

\newpage

\end{document}